\documentclass[a4paper,11pt]{article}

\usepackage{jcappub}

\newcommand{\lsim}{\protect\raisebox{-0.8ex}{$\:\stackrel{\textstyle <}{\sim}\:$}} 
\newcommand{\gsim}{\protect\raisebox{-0.8ex}{$\:\stackrel{\textstyle >}{\sim}\:$}}

\title{The magnitude-redshift relation in a realistic inhomogeneous universe}

\author{Ryuichiro Hada}
\author{and Toshifumi Futamase}
\affiliation{Astronomical Institute, Graduate School of Science, Tohoku University, 
\\  Sendai, Miyagi 980-8578, Japan}

\emailAdd{r.hada@astr.tohoku.ac.jp}
\emailAdd{tof@astr.tohoku.ac.jp}

\abstract{
The light rays from a source are subject to a local inhomogeneous geometry generated by inhomogeneous matter distribution as well as the existence of collapsed objects. In this paper we  investigate the effect of inhomogeneities and the existence of collapsed objects on the propagation of  light rays and evaluate changes in 
the magnitude-redshift relation from the standard relationship found in a homogeneous 
FRW universe. We give the expression of the correlation function and the variance for the perturbation of apparent magnitude, and calculate it numerically by using the non-linear matter power spectrum. We use the lognormal probability distribution function for the density contrast and spherical collapse model to 
truncate the power spectrum in order to estimate the blocking effect by collapsed objects. We find that the uncertainties 
in $\Omega_m$ is $\sim 0.02$, and that of $w$ is $\sim 0.04$.
We also discuss a possible method to extract these effects from  real data which contains intrinsic ambiguities associated with the absolute magnitude.
}


\arxivnumber{1410.2469}

\begin{document}
\maketitle
\flushbottom

\section{\label{sec1}Introduction}

The distance-redshift relation is a basic concept in  observational cosmology,  used to estimate the cosmological parameters of the universe. 
The observed relation for supernovae SNe Ia has suggested the existence of dark energy and is expected to be used to measure the detailed properties of dark energy in future galaxy surveys. This expectation is based on the fact that  the distance for a source at $z$ is determined  uniquely in a homogeneous FRW universe if the cosmological parameters are fixed. 

However, the light rays from a source are subject to a local inhomogeneous geometry generated by inhomogeneous matter distribution as well as the existence of collapsed objects. 
This fact causes two effects on the apparent magnitude. 
One is the systematic decrease of the average apparent magnitude of the observed sources. 
This results in the fact that  light rays cannot propagate 
through  collapsed objects and thus we can only observe the light rays which propagate through a region of rarefied density compared to the averaged density. 
If this effect is significant, it may cause misinterpretation of the observed results. 
This was  first reported by Zel'dovich~\cite{1964SvA.....8...13Z}, and there have been many studies since then~\cite{1969ApJ...155...89K,1972ApJ...174L.115D,1973ApJ...180L..31D,1998PhRvD..58f3501H,1998PThPh.100...79T,
2005PhRvD..71f3537B,2005ApJ...631..678H,2008PThPh.120..937Y,2013PhRvD..87l3526F}.  
The other effect is  dispersion of the observed apparent magnitude around the mean. This effect adds another uncertainty, in addition to an uncertainty in the knowledge of  absolute magnitude, for the  determination of  cosmological parameters 
through the observation of  standard candles such as supernovae SNe  Ia. This effect also has been investigated previously~\cite{1999MNRAS.305..746M,2006PhRvD..73b3523B,2010MNRAS.405..535J,2013JCAP...06..002B,2014MNRAS.442.2659F}.
In view of planned  future surveys of supernovae, which aim to precisely determine  cosmological parameters, and in particular the nature of dark energy, it is important to have an accurate and theoretical understanding of these effects.

In this paper we  investigate the effect of inhomogeneities and the existence of collapsed objects on the propagation of  light rays and evaluate changes in 
the magnitude-redshift relation from the standard relationship found in a homogeneous 
FRW universe.  The method we adopt here is to use the general  formula 
for the distance-redshift relation in a realistic inhomogeneous universe~\cite{1989PhRvD..40.2502F}. 
This relation gives the relative perturbation of the distance from the standard FRW distance and is expressed as the integral of the density contrast along the line of sight with a window function.   
As mentioned above, an inhomogeneous matter distribution causes two effects on the apparent magnitude including a decrease in the average, and dispersion. Both effects are conveniently treated in the distance formula we use because 
it accounts for the density contrast along the line of sight.   

When we estimate the average decrease in the apparent magnitude from the case of a homogeneous FRW universe (hereafter, we use "deviation" to denote this) by using the above formula, we need to {\it average matter-density except for collapsed objects}. This problem was studied by Okamura and Futamase~\cite{2009PThPh.122..511O}. 
In previous work, they integrate the mass function of the collapsed object proposed by Sheth and Tormen~\cite{1999MNRAS.308..119S} over a certain mass range, that is, calculate the total density of the collapsed objects, and estimate the average of matter-density except for collapsed objects by subtracting it from the average density. 
In this paper we employ, for the purpose of averaging, the lognormal probability distribution function for the density contrast and the spherical collapse model to identify the critical density at which collapsed objects begin to be formed.
The dispersion may be then calculated by using the power spectrum.  
We use a nonlinear matter power spectrum derived by Reg PT (2-loop level), which approximates the N-body result well up to $k \simeq 1$ even at small $z$.

Of course, there are errors in the apparent magnitude caused by the observational conditions and by a poor understanding of the absolute magnitude, and these errors 
are estimated to  be roughly $0.2$~\cite{1999ApJ...517..565P}, which will be larger in general than the effects  considered here, but it is still worthwhile to investigate the effects of inhomogeneities.
This is because there is hope that  progress in the theoretical understanding of supernovae as well as an increase in the number of  observed supernovae will decrease  ambiguities in the determination of the absolute and apparent magnitudes, respectively, in the future. The effects considered in this paper contain important  information on  structure formation and cosmological parameters. 
If these effects are identified,  there is a possibility to obtain such information. Although we do not consider this possibility,  
we discuss  one method to identify the effects.  

The paper is organized as follows. In Section~\ref{sec2}, we first consider  deviation of the distance in a realistic inhomogeneous universe from the FRW metric and the change in apparent magnitude emerges accordingly. We then present the perturbation of the apparent magnitude in terms of the density contrast of matter. In Section~\ref{sec3}, we first estimate the deviation of the apparent magnitude depending on the mass scale within which the density contrast is smoothed, by using the critical value of the density contrast at which collapsed objects are formed, and the lognormal PDF. Then we  account for the probability  that light rays reach us for each mass scale which the light rays have {\it felt}, and predict a plausible deviation of apparent magnitude. In Section~\ref{sec4}, we introduce the predicted non-linear power spectrum for matter perturbation, then, compute numerically the variance for the perturbation of apparent magnitude from the correlation function. In Section~\ref{sec5}, we estimate, in the distance-redshift relation, the uncertainties of the cosmological parameters, $\Omega_m$ and $w \ (p_{\Lambda}=w \rho_{\Lambda})$. Finally, Section~\ref{sec6} is devoted to conclusions and discussion.

\section{\label{sec2}Perturbation of Apparent Magnitude}

In a homogeneous flat FRW universe, the angular diameter distance and the luminosity distance are defined as follows:

\begin{eqnarray}
	d^{\rm{FRW}}_{\rm{A}}(z)
		&=&\frac{\chi(z)}{1+z}, \\
	d^{\rm{FRW}}_{\rm{L}}(z)
		&=&(1+z)\chi(z),
\end{eqnarray}

respectively, where $\chi(z)$ is the comoving distance, which is 

\begin{eqnarray}
	\chi(z)
		=\int_0^z \frac{\ dz'}{H(z')}.
\end{eqnarray}

We see from the above expression that a value is decided uniquely if the cosmological parameters are provided. In what follows, we consider how angular diameter (or luminosity) distance or apparent magnitude are modified in a realistic inhomogeneous flat\footnote{In this paper, we consider a {\it flat} universe only.} universe.

In this paper we use the distance-redshift relation in a realistic inhomogeneous universe~\cite{1989PhRvD..40.2502F,2009PThPh.122..511O},

\begin{eqnarray}
	\delta_{d}(z_s,\hat{\bf{n}})
  		&\equiv& \frac{\delta d^{\rm{FRW}}_{\rm{A}}(z_s,\hat{\bf{n}})}{d^{\rm{FRW}}_{\rm{A}}(z_s)}
		= \frac{\delta d^{\rm{FRW}}_{\rm{L}}(z_s,\hat{\bf{n}})}{d^{\rm{FRW}}_{\rm{L}}(z_s)}
  		\nonumber \\
		&=&{\bf{v}}_s \cdot \hat{\bf{n}}-\frac{1}{\chi_s}\left[\frac{1}{aH}\right]_s({\bf{v}}_s \cdot \hat{\bf{n}}-{\bf{v}}_o \cdot \hat{\bf{n}}) 
		\nonumber \\
		& &-\int_{0}^{\chi_{s}} d\chi\frac{(\chi_s-\chi)\chi}{\chi_s}\left(4 \pi G a^2 \delta \rho_m(z,\hat{\bf{n}})+\tilde{\sigma}^2\right), \label{eq.d-z}
		\nonumber \\
\end{eqnarray}

where $\hat{\bf{n}}$ is the source direction, $\chi_s \equiv \chi(z_s)$ is the source comoving distance, ${\bf{v}}_s$ and  ${\bf{v}}_o$ are the source and observer peculiar velocities respectively, $\delta \rho_m$ is the perturbation of non-relativistic matter, and $\tilde{\sigma}^2$ represents the squared shear of the bundle of light rays. The second line corresponds to the Doppler term, which has a contribution from the local peculiar velocity changing the redshift of sources relative to the observer and the solid angle of the observer. The third line corresponds to the lensing term, which has a contribution from the  inhomogeneity of the line-of-sight matter distribution. In this paper, we account for the contribution from the perturbation of non-relativistic matter {\it only} to estimate the deviation of the distance, and don't consider the Doppler term or the effect of shear in the equation above.\footnote{The sher corresponds to second order differential of gravitational potential, therefore, it seems that the effect is not very critical.}

The magnitude-redshift relation is then 
derived from the distance-redshift relation as follows:

\begin{eqnarray}
	m(z_s)
		&=&5\log_{10} d_{\rm{L}}(z_s)+M 
		\nonumber \\
		&=&\frac{5}{\ln10}\ln d_{\rm{L}}(z_s)+M, \label{eq.m-db}
\end{eqnarray}

where $M$ is the absolute magnitude. Hence, the difference in the apparent magnitude due to the variation of the (luminosity) distance $d_{\rm{L}} \rightarrow d_{\rm{L}}+\delta d_{\rm{L}}$ is written, in terms of $\delta_{d}(z_s,\hat{\bf{n}})$ in Eq.~(\ref{eq.d-z}), as follows:

\begin{eqnarray}
	\delta m(z_s,\hat{\bf{n}})
		&=&\frac{5}{\ln10}\ln(1+\delta_{d}(z_s,\hat{\bf{n}})) 
		\nonumber \\
		&=&\frac{5}{\ln10}\delta_{d}(z_s,\hat{\bf{n}}). \label{eq.m-d}
\end{eqnarray}

When writing the second equality, we neglect the second or upper order term in $\delta_{d}$ based on the assumption $\delta_{d} \ll 1$. We will see the validity of this assumption in Sec.~\ref{sec5}.
 
Thus, we obtain the following expression for the variation of the apparent magnitude. 

\begin{eqnarray}
	\lefteqn{	
		\delta m(z_s,\hat{\bf{n}})}
		\nonumber \\
		& &=-\frac{5}{\ln10}\int_{0}^{\chi_{s}} d\chi\frac{(\chi_s-\chi)\chi}{\chi_s}\ 4 \pi G a^2 \delta \rho_m(z,\hat{\bf{n}}) \label{eq.m-z_o} \\
		& &=-\frac{15H_0^2\Omega_{m0}}{2\ln10}\int_{0}^{\chi_{s}} d\chi\frac{(\chi_s-\chi)\chi}{\chi_s}(1+z)\delta_m(z,\hat{\bf{n}}),\label{eq.m-z}
\end{eqnarray}

where $\delta_m$ is the relative perturbation of non-relativistic matter.
From this equation, we see that the {\it perturbation} of the apparent magnitude reduces to the perturbation of non-relativistic matter in an inhomogeneous flat universe.

\section{\label{sec3}Deviation of the Apparent Magnitude}

Hereafter, using the above relations, Eq.~(\ref{eq.m-z_o}) and (\ref{eq.m-z}), we consider the quantities caused by the inhomogeneity of the universe and related to actual observables. First, in this section, we estimate  {\it deviation} of the apparent magnitude from that in a homogeneous flat FRW universe. The deviation at each $z_s$ as a statistical quantity, $\Delta m(z_s)$, is obtained by averaging $\delta m(z_s,\hat{\bf{n}})$ over the apparent magnitudes of many sources, SNe Ia etc., at $z_s$ in various directions $\hat{\bf{n}}$: $\Delta m(z_s) \equiv \langle \delta m(z_s,\hat{\bf{n}}) \rangle$. The average of the perturbation of non-relativistic matter is obviously zero: $\langle \delta \rho_m(z,\hat{\bf{n}}) \rangle = 0$. As long as we  deal with the matter-energy density $\rho_{m} \equiv \bar{\rho}_{m}+\delta \rho_m=\bar{\rho}_m(1+\delta_m)$ as a continuous ideal fluid, it follows from Eq.~(\ref{eq.m-z_o}) that there is no deviation: $\langle \delta m(z_s,\hat{\bf{n}}) \rangle = 0$. However, in fact, since collapsed objects have been formed until the present, we need to recognize that the perturbation of matter-energy density is composed of a {\it fluid} part and a  {\it collapsed objects} part:

\begin{eqnarray}
	\delta \rho_m&=&\delta \rho_{m \ ({\rm fluid})}+\delta \rho_{m\ ({\rm coll})}, \\
		\delta_m&=&\delta_{m \ ({\rm fluid})}+\delta_{m\ ({\rm coll})},
\end{eqnarray}
where
\begin{eqnarray}
	\delta_{m \ ({\rm fluid})}=\delta \rho_{m \ ({\rm fluid})}/\bar{\rho}_{m}, \ \ \delta_{m\ ({\rm coll})}=\delta \rho_{m \ ({\rm coll})}/\bar{\rho}_{m}. 
	\nonumber
\end{eqnarray}

In this situation, if there are collapsed objects between an observer and a source observed by the observer, in other words, if there are density perturbations corresponding to $\delta \rho_{m\ ({\rm coll})}(z,\hat{\bf{n}})$ somewhere in the range $0 \sim z_s$ on the right hand side of Eq.~(\ref{eq.m-z_o}), the light rays from the source at $z_s$ cannot propagate to the observer. Therefore, to estimate  deviation of the apparent magnitude for sources {\it actually} observed, we have to take into account  the contribution of the fluid part {\it only} to the density contrast. Accordingly, we define the deviation $\Delta m(z_s)$ as

\begin{eqnarray}
	\Delta m(z_s)&\equiv&\langle \delta m(z_s,\hat{\bf{n}}) \rangle
		\nonumber \\
		&=&-\frac{5}{\ln10}\int_{0}^{\chi_{s}} d\chi\frac{(\chi_s-\chi)\chi}{\chi_s}\ 4 \pi G a^2 \langle \delta \rho_{m\ ({\rm fluid})}(z,\hat{\bf{n}}) \rangle \label{eq.m-z_d_o} \\
		&=&-\frac{15H_0^2\Omega_{m0}}{2\ln10}\int_{0}^{\chi_{s}} d\chi\frac{(\chi_s-\chi)\chi}{\chi_s}(1+z)\langle \delta_{m\ ({\rm fluid})}(z,\hat{\bf{n}}) \rangle.\label{eq.m-z_d}
\end{eqnarray}

In what follows, we consider how the average of the fluid part of the density contrast, $\langle \delta_{m\ ({\rm fluid})}(z,\hat{\bf{n}}) \rangle$ is estimated. We shall also simplify the notation for convenience: $\delta_{m}(z,\hat{\bf{n}})=\delta(z,\hat{\bf{n}})$.

\subsection{\label{sec3_A}The Critical Value of the Density Contrast}

We introduce the critical value of the density contrast where the fluid part changes to the collapsed objects part by using a {\it spherical collapse model}. In the spherical collapse model, the following property holds for non-relativistic matter: if in the {\it linearized theory} the smoothed density contrast in a ball of the present size $R$ exceed a critical value, $\delta_{{\rm lin},c}(z)$, the ball has just collapsed~\cite{Gorbunov2011b,Loeb2013}. We use the subscripts ``lin'' and ``nl'' to distinguish the variables corresponding to the primordial (linear) and the evolved (nonlinear) density fields, respectively. Moreover, the smoothed density contrast of a ball of the present size $R$, $\delta({\bf{x}};R)$ is related to the {\it unsmoothed} density contrast, $\delta({\bf{x}})$ as 

\begin{eqnarray}
	\delta({\bf{x}};R)
		=\int d^3 y \ \delta({\bf{x}}+{\bf{y}})\ W_{R}({\bf{y}}), 
\end{eqnarray}

where ${\bf{x}}$, ${\bf{y}}$ are the coordinates defined in the comoving flame, related to the above $\hat{\bf{n}}$ with the relation, ${\bf{x}}=\chi(z)~\hat{\bf{n}}$,\footnote{In this position, hereafter, we often use $\delta(z,\hat{\bf{n}})$ to denote $\delta({\bf{x}})$.} and $W_{R}({\bf{y}})$ is the top hat window function  

\begin{eqnarray}
	W_{R}({\bf{y}})
		=\frac{3}{4\pi}\frac{1}{R^3}\theta(R-\left|{\bf{y}}\right|), 
\end{eqnarray}

($\theta$: step function). With this choice, the relation between the size R and the mass of the collapsed object is standard,

\begin{eqnarray}
	M(R)=\frac{4\pi}{3}R^3 \bar{\rho}_{m ,0},
\end{eqnarray}

where $\bar{\rho}_{m,0}$ is the present  energy density of non-relativistic matter. The critical value $\delta_{{\rm lin},c}(z)$ is obtained analytically in the cosmological model without dark energy: $\delta_{{\rm lin},c}(z)=1.686$. 
In the universe with dark energy it becomes $\delta_{{\rm lin},c}(z)$ time-dependent~\cite{2009PThPh.122..511O,2003MNRAS.344..149H}:

\begin{eqnarray}
	\delta_{{\rm lin},c}(z) 
		=1.686 \ \Omega_{m}^{0.0055} (z), \label{L_crit}
\end{eqnarray}

where $\Omega_{m}(z)$ is the time-dependent cosmological parameter for the matter component. \par
Hence, a region of size R centered at the point ${\bf{x}}$ has collapsed by the redshift $z$ providing that the following condition holds,

\begin{eqnarray}
	\delta_{{\rm lin}}({\bf{x}};R)
		=\delta_{{\rm lin}}(z,\hat{\bf{n}};R)
		\geq \delta_{{\rm lin},c}(z).  \label{coll_cri}
\end{eqnarray}

We note that the quantity on the left hand side of Eq.~(\ref{coll_cri}) does not have direct physical significance. The function $\delta_{{\rm lin}}({\bf{x}};R)$  {\it would be} the  matter density contrast {\it if} the linear theory were correct all the way until redshift $z$. In reality, the density contrast in the center of a collapsing region greatly exceeds $\delta_{{\rm lin},c}(z)$ at the time when the linear theory gives $\delta_{{\rm lin}}(z,\hat{\bf{n}};R)=\delta_{{\rm lin},c}(z)$, that is, the linear theory has broken down before that.

\subsection{\label{sec3_B}Probability Density Function}

Next, we consider a probability density function (PDF) of the density contrast, which will be used, in Sec.~\ref{sec3_C}, to compute $\langle \delta_{m\ ({\rm fluid})}(z,\hat{\bf{n}}) \rangle$.  
In this paper, we use the {\it lognormal PDF} which has been shown to give an accurate 
fit with the density field predicted by a series of cosmological $N$-body simulations in three CDM models (standard, lambda, and open CDM)~\cite{2001ApJ...561...22K}. For simplicity we use $\delta$ to denote $\delta({\bf{x}};R)$ unless otherwise stated. The lognormal PDF of a field $\delta$ smoothed over $R$ is defined as

\begin{eqnarray}
	p_{LN}(z,\delta;R) 
		=\frac{1}{(2\pi \sigma_1^2)^{1/2}}\exp \left\{-\frac{[\ln(1+\delta)+\sigma_1^2/2]^2}{2\sigma_1^2}\right\}\frac{1}{1+\delta}. \label{P_LN}
\end{eqnarray} 

Here, $\sigma_1$ depends on the smoothing scale $R$ and is given by 

\begin{eqnarray}
	\sigma_1^2(z;R) 
		&=&\ln[1+\sigma_{\rm nl}^2(z;R)], \label{sig_1} \\
		\sigma_{\rm nl}^2(z;R)
		&\equiv& \frac{1}{2\pi^2} \int_{0}^{\infty}dk \ k^2 \ \tilde{W}_R^2(k)P_{\rm nl}(z,k),
\end{eqnarray}

where $\tilde{W}_R(k)=3(\sin kR -kR \cos kR)/(kR)^3$ is the Fourier transformation of $W_{R}({\bf{x}})$ and $P_{\rm nl}(z,k)$ is the power spectrum at a redshift $z$.\footnote{In fact, if we assume the random variable $(1+\delta)$ obeys a lognormal PDF, $\sigma_1$ (that is, the logmarmal PDF) is decided uniquely by Eq.~(\ref{P_LN}) and (\ref{sig_1}) from the condition that the mean and variance of the density contrast $\delta$ are zero and $\sigma_{\rm nl}$, respectively~\cite{1991MNRAS.248....1C}.} \par

It is known that there is a one-to-one correspondence between the nonlinear density contrast field $\delta$ whose distribution is given by the above lognormal PDF and the linear random Gaussian~\cite{1991MNRAS.248....1C}. We define a linear density contrast field $g\equiv g({\bf{x}};R)$ smoothed over $R$ obeying the Gaussian PDF,

\begin{eqnarray}
	p_{G}(z,g;R) 
		=\frac{1}{(2\pi \sigma_{\rm lin}^2)^{1/2}}\exp \left(-\frac{g^2}{2 \sigma_{\rm lin}^2}\right) , \label{P_G}
\end{eqnarray} 

where the variance $\sigma_{\rm lin}$ is computed from its linear power spectrum

\begin{eqnarray}
	\sigma_{\rm lin}^2(z;R)
		\equiv \frac{1}{2\pi^2} \int_{0}^{\infty}dk \ k^2 \ \tilde{W}_R^2(k)P_{\rm lin}(z,k).
\end{eqnarray}

Then, the relation between $\delta$ and $g$ is written as

\begin{eqnarray}
	1+\delta
		=\exp \left(\frac{\sigma_1}{\sigma_{\rm lin}}\ g -\frac{\sigma_1^2}{2}\right). \label{trans}
\end{eqnarray}

We can verify the above relation by the fact that the PDF for $\delta$ is simply given by $(dg/d\delta)p_G(g)$, which reduces to Eq.~(\ref{P_LN}).  \par

A few comments concerning the lognormal PDF are in order. First, the  density contrast field $\delta$ obeying the lognormal PDF always has $\delta > -1$ from Eq.~(\ref{trans}). Accordingly, the condition $\rho > 0$ holds automatically, as expected. Second, at an early time when the linear theory is correct, that is, $g \ll 1$,  it follows from Eq.~(\ref{sig_1}), up to the linear order, that

\begin{eqnarray}
	\mbox{ RHS of Eq. (\ref{trans})}
		\simeq 1+g. 
\end{eqnarray}

(Here we assume $\sigma_{\rm nl}=\sigma_{\rm lin} \ll g$.) We see that the lognormal PDF Eq.~(\ref{P_LN}) is close to the Gaussian at the early time. From these properties, we see that the lognormal PDF indeed satisfies some physical requests, however, there is no physical reason to believe that the density contrast field $\delta$ evolved from $g$.\footnote{There is a simple model based on the {\it dark halo approach} which explains the shape of the non-Gaussian tails of the lognormal PDF~\cite{2003MNRAS.339..495T}.} Nevertheless, the lognormal PDF has been shown to provide a good fit to the simulation data empirically.

\subsection{\label{sec3_C}The Mean Density Averaged Over a Mass Scale and the Deviation of the Apparent Magnitude}

We will now compute the average of the fluid part of the density contrast in Eq.~(\ref{eq.m-z_d}), $\langle \delta_{m\ ({\rm fluid})}(z,\hat{\bf{n}}) \rangle$. To summarize the results of Sec.~\ref{sec3_A} and Sec.~\ref{sec3_B}, we know that the smoothed density contrast $\delta=\delta({\bf{x}};R)=\delta(z,\hat{\bf{n}};R)$ obeys the lognormal PDF Eq.~(\ref{P_LN}) throughout the evolution. At an early time, $\delta \ll 1$, the PDF is a Gaussian distribution Eq.~(\ref{P_G}), thereafter, as $\delta$ grows larger, the peak gradually shifts to the side $\delta<0$ and the tail of the distribution on the side $\delta>0$ extends in the direction $\delta \to \infty$. Then, when the density contrast reaches $\delta \simeq 1$, various scale modes begin to interact with each other, and  growth enters the non-linear regime. If after that, the density contrast evolves further  in the linear theory (not actually), it would collapse at $z_{\rm coll}$ such that $\delta_{{\rm lin}}(z_{\rm coll},\hat{\bf{n}};R)= \delta_{{\rm lin},c}(z_{\rm coll})$.  \par

Here, we assume that a light ray from a source has propagated all the way through the density field smoothed within a ball of $R$ (or mass scale $M(R)$). Then, we can replace the density contrast (including the collapsed objects part) by the smoothed one averaged over a scale $R$: we can use, in Eq.~(\ref{eq.m-z}), $\delta(z,\hat{\bf{n}};R)$ as $\delta_{m}(z,\hat{\bf{n}})$. In the linear theory, $g<\delta_{{\rm lin},c}$ and $g>\delta_{{\rm lin},c}$ corresponds to the fluid  and  collapsed objects parts, respectively. Therefore, it seems possible to obtain an average of the fluid part by averaging the smoothed linear density contrast $g(z,\hat{\bf{n}};R)$ with the linear random Gaussian $p_{G}(z,g;R)$ in the range from $-1$ to $\delta_{{\rm lin},c}$. However, when the linear density contrast grows to $g \gsim 1$, the PDF $p_{G}(z,g;R)$ doesn't satisfy the condition $g>-1$. To prevent that, we use  the lognormal PDF $p_{LN}(z,g;R)$ computed with the replacement $\sigma_{\rm nl} \to \sigma_{\rm lin}$: $p_{LN}^{\rm lin}(z,g;R)$ instead of $p_{G}(z,g;R)$. Finally, based on the above discussion, we can estimate $\langle \delta_{m\ ({\rm fluid})}(z,\hat{\bf{n}}) \rangle$ as a quantity depending on the smoothing mass scale $M(R)$:
\begin{eqnarray}
	\langle \delta_{m\ ({\rm fluid})}(z,\hat{\bf{n}}) \rangle_M
		=\int_{-1}^{\delta_{{\rm lin},c}(z)} dg \  g \ p_{LN}^{\rm lin}(z,g;R) \label{del_ave}.
\end{eqnarray}

From the above equation, we find that the right hand side of Eq.~(\ref{eq.m-z_d}) is dependent on the smoothing mass scale $M(R)$, accordingly, we have to recognize the deviation $\Delta m(z_s)$ to be the quantity decided by $M$. Thus, we redefine this deviation on the left hand side of Eq.~(\ref{eq.m-z_d}) as $\Delta m(z_s;M)$. \par

Finally, from Eq.~(\ref{eq.m-z_d}), (\ref{P_LN}), and (\ref{del_ave}), we can compute the deviation of the apparent magnitude from a homogeneous flat FRW universe provided {\it only} that the cosmological parameters are given.\footnote{In computing $p_{LN}^{\rm lin}$, we need also the initial value of {\it linear} power spectrum $P_{\rm lin}(k)$, however, if the cosmological parameters are given, it can be obtained by {\it CAMB Web Interface} : \url{http://lambda.gsfc.nasa.gov/toolbox/tb\_camb\_form.cfm}} In what follows, we use the cosmological parameters presented by the {\it Wilkinson Microwave Anisotropy Probe} five year release~\cite{2009ApJS..180..330K}. \par

\begin{figure}
\begin{center}
\includegraphics[clip,width=9cm,angle=270]{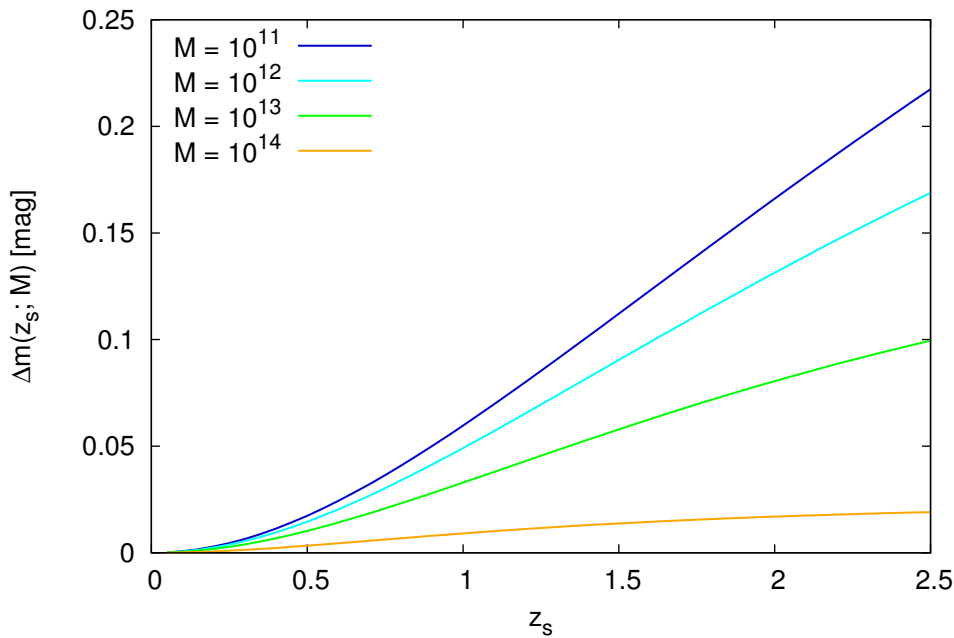}
\end{center}
\caption{\label{fig:Dm(z_s;M)} Deviation of the apparent magnitude from a homogeneous flat FRW universe at each redshift $z_s$ corresponding to sources observed for various smoothing mass scales: $10^{11}M_{\odot}$(blue), $10^{12}M_{\odot}$(cyan), $10^{13}M_{\odot}$(green), and $10^{14}M_{\odot}$(orange).}
\end{figure}

In Fig.~\ref{fig:Dm(z_s;M)}, we show the relation between $\Delta m(z_s;M)$ and $z_s$ for various smoothing scales: $(M(R)\ [M_{\odot}],\ R\ \mbox{[Mpc/h]})=(10^{11},0.61)$, $(10^{12},1.30)$, $(10^{13},2.81)$, and $(10^{14},6.05)$ are shown in blue, cyan, green, and orange lines, respectively. First, we see that on all mass scales, the apparent magnitude of a further source is shifted largely from the homogeneous universe to the side $\Delta m(z_s;M) > 0$. This fact is explained as follows: the light ray from a further source is more easily  blocked  by collapsed objects, hence, a light ray that is actually observed  has propagated through a low density area ($\rho < \bar{\rho}$, that is, $\delta < 0$) for a {\it longer} time. Finally, we observe the source darker by demagnification in gravitational lensing (or by Eq.~(\ref{eq.m-z_o})). This figure also shows that the smaller the smoothing mass scale $M(R)$ becomes, the larger the deviation becomes. When we think of light rays from sources at the same $z_s$, the density contrast smoothed in a ball of smaller $R$ (or smaller $M(R)$) is more easily  collapsed. Thus, a light ray observed actually has propagated through a {\it lower} density area. It follows by the same reasoning as above, that the deviation $\Delta m(z_s;M) (> 0)$ becomes larger as the smoothing mass scale is smaller.

\subsection{\label{sec3_D}The Observed Deviation}  

In Sec.~\ref{sec3_C}, we  assumed that light rays are subjected to a matter-density distribution smoothed within a certain mass scale along the way, and obtained the deviation $\Delta m(z_s;M)$ depending on the mass scale. In reality,  light rays which feel the density distribution smoothed on a smaller mass scale are more easily blocked  by collapsed objects, and harder to propagate to the observer since $\sigma_{\rm lin}(z;M(R)) \to \infty$ as $M \to 0$, that is, the  small mass scale is easy to collapse. Thus, we need to take account of the probability that a light ray from $z_s$ can propagate to the observer at each smoothing mass scale. In what follows, we estimate the probability by using the idea of {\it transmittance}, then, obtain the deviation of apparent magnitude averaged over the smoothing mass scale. \par

We consider spherical collapse in the Press-Schechter model~\cite{Gorbunov2011b,Loeb2013,2001MNRAS.323....1S}. Letting \\ $(dn(z;M)/dM)dM$ be the (physical) number density of halos of mass between $M$ and $M+dM$ at $z$, and have 

\begin{eqnarray}
	\frac{dn(z;M)}{dM}=\sqrt{\frac{2}{\pi}}\ \frac{\bar{\rho}_m}{M}\ \frac{d\nu}{dM}\ \exp \left(-\frac{\nu^2}{2}\right), \label{P_S}
\end{eqnarray}

where $\bar{\rho}_m$ is the background matter-density, $\nu \equiv \delta_{{\rm lin},c}(z)/\sigma_{\rm lin}(z;M)$. Hence, we obtain the number density of objects with mass exceeding a given value at $z$,

\begin{eqnarray}
	n(z;>M)=\int^{\infty}_{M} \frac{dn(z;M')}{dM'} dM'.
\end{eqnarray}

Accordingly, the photon mean free path at $z$ for the light rays which feel  the matter-density distribution smoothed within a mass scale $M$ is estimated as follows,

\begin{eqnarray}
	\lambda_{\gamma}(z;M) 
		\simeq \frac{1}{A_{\rm coll}(z;M)\ n(z;>M)},
\end{eqnarray}

where

\begin{eqnarray}
	A_{\rm coll}(z;M)
		= \pi \left(\frac{R_{\rm vir}(M)}{1+z}\right)^2, 
		\ \ M \equiv \frac{4\pi}{3}R_{\rm vir}^3(M)\ \bar{\rho}_{m,0}(1+\delta_{\rm vir})
\end{eqnarray}

is the cross section of photons blocked  by collapsed objects. Here $(1+\delta_{\rm vir})$ is the density of the virialized object relative to the critical density, we set this value as $\delta_{\rm vir}=177$, which is derived for a universe with $\Omega_{m}=1$~\cite{Loeb2013}, for this estimate. There are several comments to make. First, the above photon mean free path,  $\lambda_{\gamma}(z;M)$, is underestimated since we use $n(z;>M)$ for the number density of collapsed objects at mass scale $M$.\footnote{However, in fact, the number density of halos $dn(z;M)/dM$ decreases exponentially and the halos at mass scale $M$ are dominant, thus, this estimate is  reasonable.} Second, we have assumed, for simplification, that the cross section of photons blocked  by a structure is just given by the area of the projection on the sky of the structure (that is, its dark matter distribution). However, this representation is not extremely accurate since, for instance, the dark matter halo of a galaxy extends well beyond the stellar distribution of the galaxy.

Hence, we obtain the probability that light rays from source at $z_s$ is not blocked  by collapsed objects up to the observer, that is, the transmittance of the light ray:

\begin{eqnarray}
	T(z_s;M)
		=\exp[-\tau(z_s;M)],
\end{eqnarray}

where

\begin{eqnarray}
	\tau(z_s;M)
		&=&\int^{t(0)}_{t(z_s)}\frac{dt'}{\lambda_{\gamma}(t';M)}
		\nonumber \\
		&=&\int^{z_s}_{0}\frac{dz'}{\lambda_{\gamma}(z';M)(1+z')H(z')}
\end{eqnarray}

is the optical depth. 

\begin{figure}
\begin{center}
\includegraphics[clip,width=9cm,angle=270]{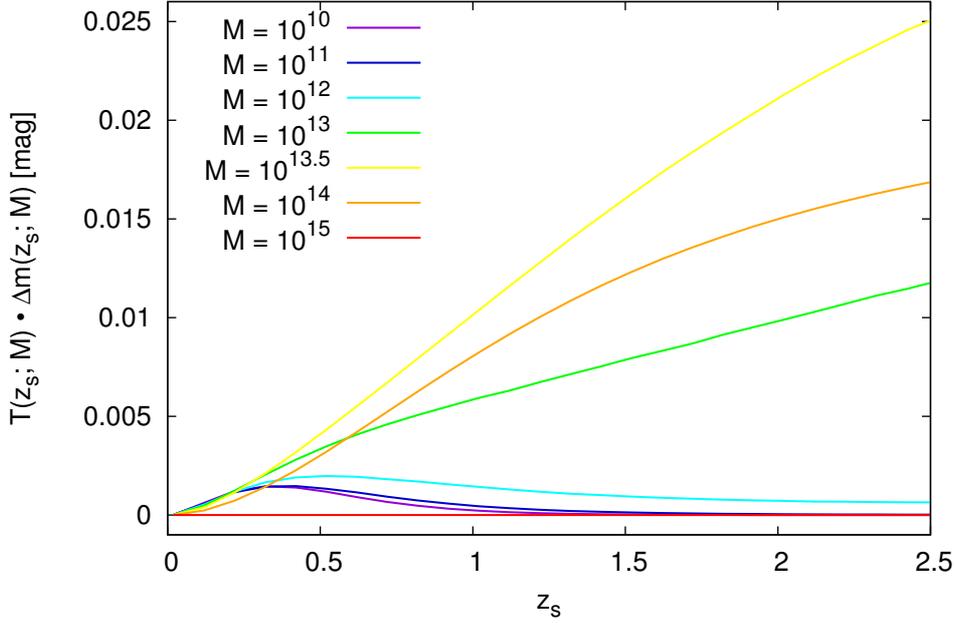}
\end{center}
\caption{\label{fig:T(z_s;M)Dm(z_s;M)} Deviation of the apparent magnitude, multiplied by the transmittance, from  a homogeneous flat FRW universe for various smoothing mass scales: $10^{10}M_{\odot}$(purple), $10^{11}M_{\odot}$(blue), $10^{12}M_{\odot}$(cyan), $10^{13}M_{\odot}$(green), $10^{13.5}M_{\odot}$(yellow), $10^{14}M_{\odot}$(orange), and $10^{15}M_{\odot}$(red).}
\end{figure}

In Fig.~\ref{fig:T(z_s;M)Dm(z_s;M)}, we show the relation between $T(z_s;M) \cdot \Delta m(z_s;M)$ and $z_s$ at various smoothing scales, including $M(R)\ [M_{\odot}]=10^{10}$, $10^{11}$, $10^{12}$, $10^{13}$, $10^{13.5}$, $10^{14}$, and $10^{15}$ are drawn with purple, blue, cyan, green, yellow, orange, and red lines, respectively. We can recognize this value, $\Delta m(z_s;M)$ weighted by $T(z_s;M)$, as the contribution of each smoothing mass scale to the net (average) deviation. From this figure, we see that the deviation becomes larger as the mass scale increases from $10^{10}M_{\odot}$. This result is opposite that discussed in Sec.~\ref{sec3_C},   due to the fact that the transmittance on a  small mass scale is so low that it suppresses the effect of the deviation $\Delta m(z_s;M)$. However, for mass scales larger  than $10^{13.5}M_{\odot}$, the deviation weighted by the transmittance decreases rapidly because, as the mass scale increases, the contribution made by a smaller deviation is more dominant than the contribution made by a larger  transmittance. \par

Based on these results, we estimate the deviation of the apparent magnitude averaged over the smoothing mass scale. We set the (normalized) probability that a light ray observed from $z_s$ has felt the matter-density distribution smoothed over the mass scale between $M$ and $M+dM$ along the way, as $w(z_s;M)dM$. This probability is expected to be proportional to the above transmittance:

\begin{eqnarray}
	w(z_s;M) \propto T(z_s;M).
\end{eqnarray}   
    
Then, we can compute the averaged deviation for the light ray from $z_s$ as follows

\begin{eqnarray}
	\Delta m(z_s)_{\rm ave}
		&\equiv& \int_{M_{\rm mim}}^{M_{\rm max}}\Delta m(z_s;M')\ w(z_s;M') \ dM' \\
		&=&\frac{\int_{M_{\rm mim}}^{M_{\rm max}}\Delta m(z_s;M')\ T(z_s;M') \ dM'}{\int_{M_{\rm mim}}^{M_{\rm max}}T(z_s;M') \ dM'}. \label{Dm_ave} 
\end{eqnarray}

We note that $\Delta m(z_s)_{\rm ave}$ represents the {\it minimum} deviation from a homogeneous universe since the photon mean free path,  $\lambda_{\gamma}(z;M)$, is underestimated. The result of this averaged deviation is discussed in Sec.~\ref{sec5}.

\section{\label{sec4}Variance of the Observed Apparent Magnitude}

We will now estimate the {\it variance} of the observed apparent magnitude. As we  mentioned in Sec.~\ref{sec1}, this variance has been previously studied~\cite{1999MNRAS.305..746M,2006PhRvD..73b3523B,2010MNRAS.405..535J,2013JCAP...06..002B,2014MNRAS.442.2659F}.
This variance is the mean square of perturbations of the apparent magnitude, thus, as shown in Sec.~\ref{sec3}, we still have to recognize the density contrast as being composed of a fluid part and a collapsed objects part (this is a significant difference comparing the standpoints in the above previous studies and this study) and consider how we average them. For this purpose it is straightforward to use 
the {\it power spectrum} of matter-perturbation. We will begin by describing the {\it correlation function}.

\subsection{\label{sec4_A}Correlation Function}

 The correlation function generally represents  the average over the entire statistical ensemble of two points separated by an interval. We want to know the variance corresponding to the apparent magnitude for sources at the same $z_s$, therefore, we must average the apparent magnitudes of two sources separated by an {\it angle} over the entire sphere of $z=z_s$. Using Eq.~(\ref{eq.m-z}), we write the correlation function as follows:
 
\begin{eqnarray}	
		\langle \delta m(z_s,{\hat{\bf{n}}})\delta m(z_s,\hat{{\bf{n}}}') \rangle
		&=&\left(\frac{15H_0^2\Omega_{m0}}{2\ln10}\right)^2 \int_{0}^{\chi_{s}} d\chi_1 \frac{(\chi_s-\chi_1)\chi_1}{\chi_s}(1+z_1)
		\nonumber \\
		&\times&\int_{0}^{\chi_{s}} d\chi_2\frac{(\chi_s-\chi_2)\chi_2}{\chi_s}(1+z_2)\langle \delta_m(z_1,\hat{\bf{n}})\delta_m(z_2,\hat{\bf{n}}') \rangle
		\nonumber \\
		&\simeq& \left(\frac{15H_0^2\Omega_{m0}}{2\ln10}\right)^2 \int_{0}^{\chi_{s}} d\chi \left[\frac{(\chi_s-\chi)\chi}{\chi_s}(1+z)\right]^2
\langle \delta_m(z,\hat{\bf{n}})\delta_m(z,\hat{\bf{n}}') \rangle \label{a_cor}
		\nonumber \\
\end{eqnarray}  

where we used the well-established approximation that the correlation of line-of-sight of the density contrast can be neglected.\footnote{This assumption is based on {\it Limber's equation}~\cite{2007A&A...473..711S}.} Here, the correlation function for the matter perturbation in the above equation is, in terms of the power spectrum~\cite{Gorbunov2011b},

\begin{eqnarray}
	\langle \delta_m(z,\hat{\bf{n}})\delta_m(z,\hat{\bf{n}}') \rangle
		&=&\langle \delta_m(z,{\bf{x}})\delta_m(z,{\bf{x}}') \rangle
		\nonumber \\
		&=&\int_0^\infty \frac{k^2dk}{2\pi^2}\frac{\sin(kr)}{kr}P_{\rm nl}(z,k), \label{m_cor}
\end{eqnarray}  

where  

\begin{eqnarray}
	r=\left|{\bf{x}}-{\bf{x}}'\right|=\chi(z)\sqrt{2(1-\cos\theta)},
		\ \ \mbox{with} \ \ 
		\cos\theta
		\equiv \hat{\bf{n}}\cdot\hat{\bf{n}}'.
		\nonumber
\end{eqnarray}

Finally, from Eq.~(\ref{a_cor}) and (\ref{m_cor}), we  express the correlation function for the apparent magnitude as a function of $\theta$:

\begin{eqnarray}
	\xi_m(z_s,\theta) 
		&\equiv& \langle \delta m(z_s,{\hat{\bf{n}}})\delta m(z_s,\hat{{\bf{n}}}') \rangle
		\nonumber \\
		&=& \left(\frac{15H_0^2\Omega_{m0}}{2\ln10}\right)^2 \int_{0}^{\chi_{s}} d\chi \left[\frac{(\chi_s-\chi)\chi}{\chi_s}(1+z)\right]^2
\int_0^\infty \frac{k^2dk}{2\pi^2}\frac{\sin(kr)}{kr}P_{\rm nl}(z,k). \label{am_cor}
		\nonumber \\
\end{eqnarray}

Note that the above expression includes the contributions of the collapsed objects $\delta_{m\ ({\rm coll})}$, which should be eliminated, as shown in Sec.~\ref{sec3}.

\subsection{\label{sec4_B}Non-linear Matter Power Spectrum}

Next, we introduce {\it Extension of Reg PT (Two-loop Level)}~\cite{2013ApJ...769..106S} as the predicted non-linear power spectrum for matter perturbation, $P_{\rm nl}(z,k)$. They are represented in terms of  standard perturbation theory (SPT):

\begin{eqnarray}
	P_{\rm nl}(z,k)
		&=&D^2 P_{\rm lin}(k)+D^4 P_{\rm{1-loop}}(k)+D^6 P_{\rm{2-loop}}(k)
		\nonumber \\
		&+&2D^8\left(\Gamma^{(1)}_{\rm{1-loop}}(k)+\frac{k^2 \sigma^2_{v}}{2}\right)\left(\Gamma^{(1)}_{\rm{2-loop}}(k)+\frac{k^2 \sigma^2_{v}}{2}\Gamma^{(1)}_{\rm{1-loop}}(k)+\frac{1}{2}\left(\frac{k^2 \sigma^2_{v}}{2}\right)^2\right)P_{\rm lin}(k)
		\nonumber \\
		&+&D^8\left[P_{44a}(k)+\frac{k^2 \sigma^2_{v}}{2}P_{24}(k)+\frac{(k^2 \sigma^2_{v})^2}{4}P_{22}(k)-k^2 \sigma^2_{v}\left(\Gamma^{(1)}_{\rm{1-loop}}(k)+\frac{k^2 \sigma^2_{v}}{2}\right)^2 P_{\rm lin}(k)\right]
		\nonumber \\
		&+&D^{10}\left(\Gamma^{(1)}_{\rm{2-loop}}(k)+\frac{k^2 \sigma^2_{v}}{2}\Gamma^{(1)}_{\rm{1-loop}}(k)+\frac{1}{2}\left(\frac{k^2 \sigma^2_{v}}{2}\right)^2\right)^2 P_{\rm lin}(k). \label{P_nl}
\end{eqnarray}

where $D=D(z)$ is the linear growth factor and $\sigma^2_v \equiv \int dp P_L(p)/6\pi^2$ is the velocity dispersion of (non-relativistic) matter 
(see also~\cite{2012ApJ...760..114S} for the definitions of other functions, $P_{\rm{1-loop}}(k)$, $P_{24}(k)$, etc). To numerically compute the above power spectrum, we used the program (code {\it RegPT}) which is available on A.Taruya's homepage.\footnote{\url{http://www2.yukawa.kyoto-u.ac.jp/~atsushi.taruya/}} We note that {\it only}  cosmological parameters are  needed to calculate the above power spectrum, and this {\it Extension of Reg PT (Two-loop Level)} is roughly consistent  with $N$-body simulations up to  $k \simeq 1\ \mbox{[h/Mpc]}$ even at relatively small $z$ (see~\cite{2013ApJ...769..106S} for details).

\subsection{\label{sec4_C}Variance}

Finally, we calculate the variance of the apparent magnitude for a source at $z_s$ from the expression for the correlation function and the predicted non-linear power spectrum given above.  \par

To begin with, we remove the collapsed objects part.  Since the predicted power spectrum in Section~\ref{sec4_B} is valid for $k \lsim 1$, and the space scale range $k \lsim 1$, that is, $R \gsim 2\pi$ corresponds to collapsed objects of the mass scale $M \gsim 10^{14}M_{\odot}$, which have been hardly formed until now in the Press-Schechter model,\footnote{At present, in the case of $M\sim10^{14}M_{\odot}$, the variance is $\sigma_{\rm lin}(z;M)= \delta_c(z)/\sqrt{2}$ where  the exponent in the suppression factor of the differential spectrum Eq.~(\ref{P_S}) is equal to 1 \citep{Loeb2013,Gorbunov2011b}.} we solve this problem by setting the maximum value of the wave number for the integral in Eq.~(\ref{am_cor}) as $k_{\rm max} \equiv 1$. In this way, we can at least eliminate the collapsed objects part of the density contrast all the way from $z=0$ to $z=z_s$. In fact, at higher $z$, the maximum mass scale of the collapsed objects is smaller. Thus, the fluid part of the power spectrum contributes to the correlation function up to a larger $k$. This estimate will be the {\it minimum} value for the correlation of the apparent magnitude. However, since the magnitude of the predicted power spectrum becomes smaller as $z$ is higher, the effect does not seem to be critical. \par

From the above discussion, we can compute the variance, $\sigma_m(z_s)$, by averaging the correlation function given by Eq.~(\ref{am_cor}) over angle $\theta$: 
   
\begin{eqnarray}
	\sigma_m^2(z_s) 
		\equiv \frac{1}{\pi}\int^{\pi}_{0}d\theta \ \xi_m(z_s,\theta). \label{sig_m} 
\end{eqnarray}

A few remarks are in order at this time. First, this variance represents the {\it minimum} size for the apparent magnitude of a source at $z_s$, therefore, when we observe  the apparent magnitude of many sources at $z_s$ whose absolute magnitude can be estimated (e.g. SNe Ia), it expresses the {\it minimum} extent of dispersion of their values, $m-M$,\footnote{Since $m-M$ is decided only by the  distance to the source (see Eq.~(\ref{eq.m-db})), if the deviation of distance, $\delta_{d}(z_s,\hat{\bf{n}})$, does not exist, it is expected that the values corresponding  to many sources at $z_s$ coincide with each other.} even if we don't  account for the errors associated with their absolute magnitude or observation of their apparent magnitude. Second, it follows from Sec.~\ref{sec4_B} and Sec.~\ref{sec4_C} that the variance $\sigma_m(z_s)$ for a source at $z_s$  is obtained uniquely if the cosmological parameters are given. In what follows, we calculate $\sigma_m(z_s)$ with the cosmological parameters presented by the {\it Wilkinson Microwave Anisotropy Probe} five year release~\citep{2009ApJS..180..330K},  just as we did for the deviation in Sec.~\ref{sec3}.

\section{\label{sec5}Uncertainty of the Cosmological Parameters}

Finally, based on the results in Sec.~\ref{sec3} and Sec.~\ref{sec4}, we estimate the uncertainty of measuring cosmological parameters using the magnitude-redshift relation for SNe Ia. We focus on two parameters: $\Omega_m$, which is relative contribution of non-relativistic matter and $w$, which decides the equation of state for dark energy, $p_{\Lambda}=w \rho_{\Lambda}$. 
Based on the above discussions, we need to consider two effects, the shift from values in the actual universe and the uncertainties around the shifted values,\footnote{Strictly speaking, the variance in Sec.~\ref{sec4} must be estimated as $\langle (\delta m-\Delta m)^2 \rangle=\langle \delta m^2 \rangle-\Delta m^2$. However, due to setting the maximum value of the wave number, we didn't consider the contribution of collapsed objects, which causes the deviation in Sec.~\ref{sec3}. Thus, we can recognize $\sigma_m(z_s)$ as the minimum dispersion around the shifted value.} which exist {\it at least} on the same level with the deviation $\Delta m(z_s)_{\rm ave}$ and the variance $\sigma_m(z_s)$, respectively, even if the errors related to the analysis or observation don't exist. \par

\begin{figure}
\begin{center}
\includegraphics[clip,width=9cm,angle=270]{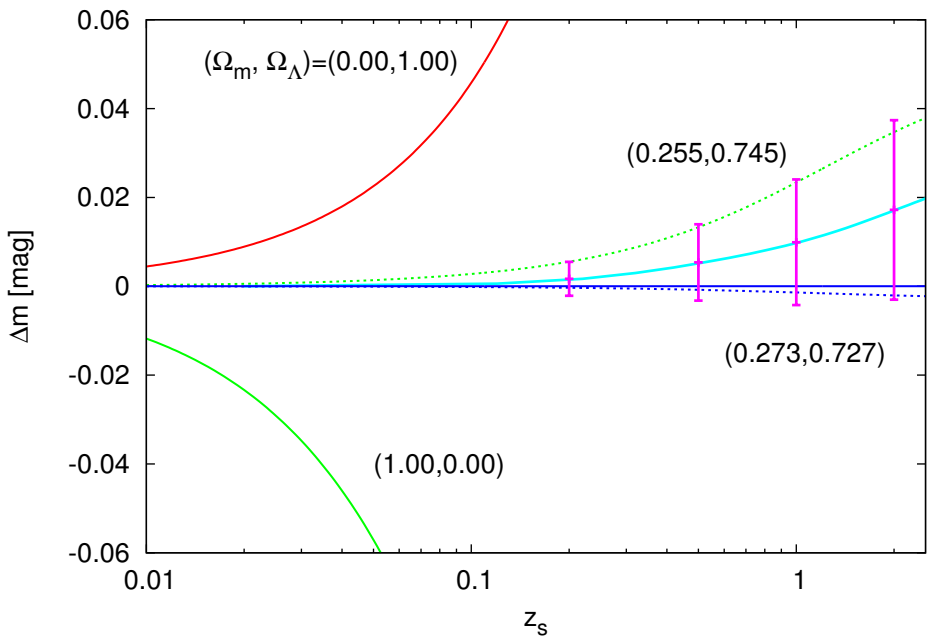}
\end{center}
\caption{\label{fig:Omega_m} Difference $\Delta m$ between $m-M$ computed using the {\it WMAP 5-year} parameters $(\Omega_m,\Omega_{\Lambda})$; $(0.272, 0.728)$ [solid blue line] and that computed by other combinations of $(\Omega_m, \Omega_{\Lambda})$; $(0.0, 1.0)$ [solid red line], $(1.0, 0.0)$ [solid green line], $(0.255, 0.745)$ [green dashed line], and $(0.273, 0.727)$ [blue dashed line]. The deviation $\Delta m(z_s)_{\rm ave}$ is shown by the cyan line, the variances $\sigma_m(z_s)$ are placed at $z_s=0.2, 0.5, 1.0, 2.0$.}
\end{figure}

\begin{figure}
\begin{center}
\includegraphics[clip,width=9cm,angle=270]{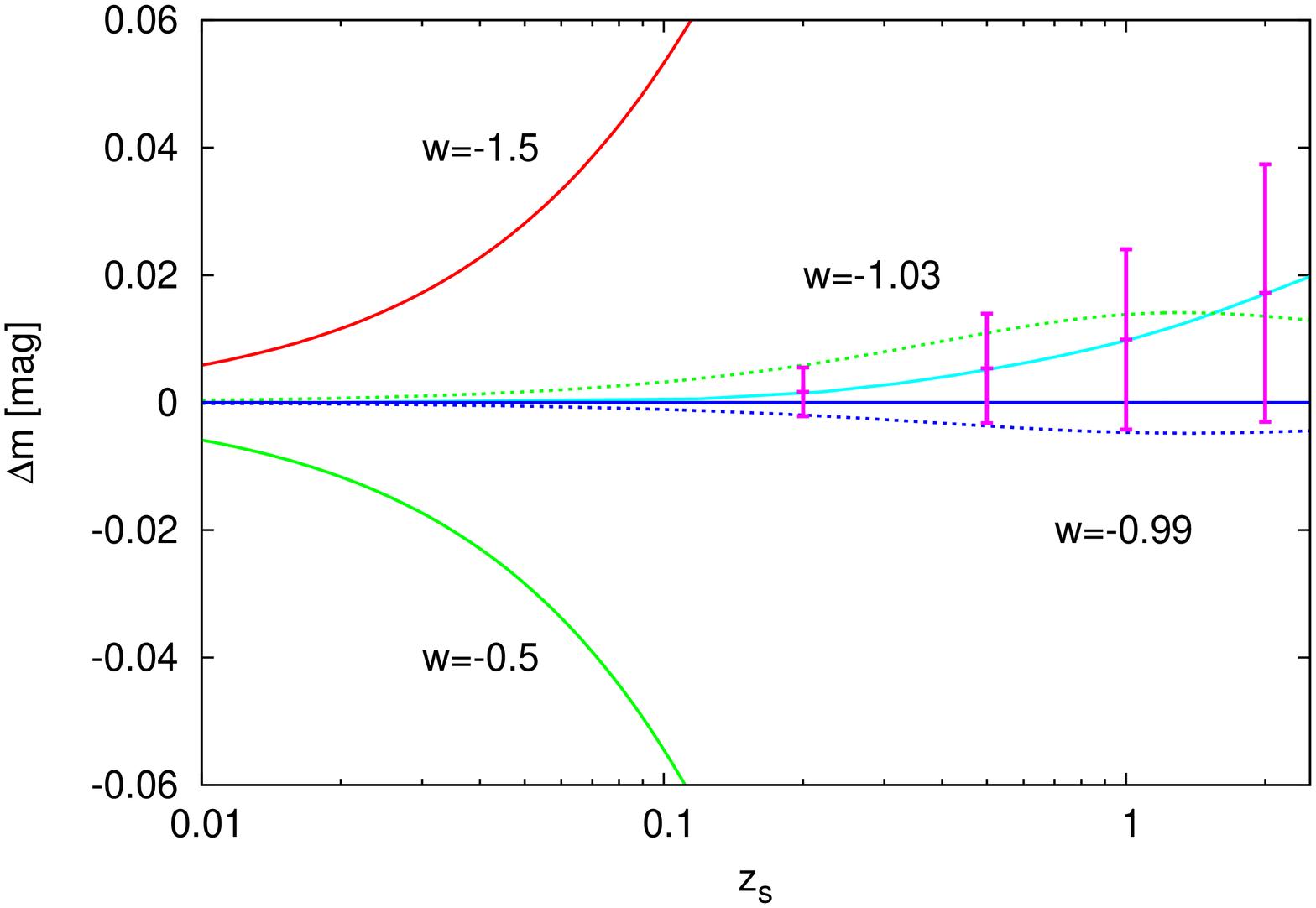}
\end{center}
\caption{\label{fig:omega} Difference $\Delta m$ between $m-M$ computed using the {\it WMAP 5-year} parameters $w=-1.0$ [solid blue line] and that computed by other values; $w=-1.5$ [solid red line], $w=-0.5$ [solid green line], $w=-1.03$ [green dashed line], and $w=-0.99$ [blue dashed line].}
\end{figure}

First, we consider $\Omega_m$. Fig.~\ref{fig:Omega_m} shows the difference $\Delta m$ between $m-M$ computed by the {\it WMAP five year} parameters $(\Omega_m,\Omega_{\Lambda})$; $(0.272, 0.728)$ [blue solid] and that computed using other combinations of $(\Omega_m, \Omega_{\Lambda})$; $(0.0, 1.0)$ [red solid], $(1.0, 0.0)$ [green solid], $(0.255, 0.745)$ [green dashed], and $(0.273, 0.727)$ [blue dashed]. In all cases, we fixed the other cosmological parameters according to the {\it WMAP five year} parameters. The deviation $\Delta m(z_s)_{\rm ave}$ is computed with $M_{\rm mim}=10^{10}M_{\odot}, M_{\rm max}=10^{15}M_{\odot}$ [cyan solid],\footnote{The validity of this setting will be discussed in Sec.~\ref{sec6}.} the variance $\sigma_m(z_s)$ [pink error bars] are placed on the deviation $\Delta m(z_s)_{\rm ave}$ at $z=0.2, 0.5, 1.0, 2.0$ so that the width of these bar equal to $2\sigma_m(z_s)$. The combinations of  parameters corresponding to blue and green dashed lines is chosen so that these graze (or  fall  inside) the $\sigma_m(z)$ bars. This figure shows that $\Delta m(z_s=2)_{\rm ave} \simeq 0.02$, that is, in terms of the relative perturbation of the distance, $\delta_{d}(z_s=2) \simeq 0.01$ (see Eq.~(\ref{eq.m-d})), which is consistent with the assumption $\delta_{d} \ll 1$ in Sec.~\ref{sec2}. We have seen, in Sec.~\ref{sec3_C}, that the deviation depending on the smoothing mass scale is always $\Delta m(z_s;M)>0$. Since the averaged deviation $\Delta m(z_s)_{\rm ave}$ is obtained by averaging this $\Delta m(z_s;M)$ weighted with the transmittance, it should be also $\Delta m(z_s)_{\rm ave}>0$. In fact, we find that this demand is satisfied by the fact that the cyan line is higher than the blue line in Fig.~\ref{fig:Omega_m}. And, we can see that  $\sigma_m(z_s)$ becomes larger  as $z$ increases. This property reflects the fact a light ray from a further source is more strongly affected by the inhomogeneity of our realistic universe, and furthermore, is intrinsically different from the errors, which don't systematically depend on $z$. Taking the fact that the deviation $\Delta m(z_s)_{\rm ave}$ and the variance $\sigma_m(z_s)$ was computed with the {\it WMAP five year} parameters taken into account, we find that if the cosmological parameters of our universe are actually the {\it WMAP five year} parameters, $\Omega_m$ (or $\Omega_{\Lambda}$) estimated in the distance-redshift relation of SNe Ia has at least the uncertainty $0.255 \sim 0.273\ ({\rm or}\ 0.727 \sim 0.745)$ since the interval between the blue and green dashed lines is roughly consistent  with $\sigma_m(z_s)$ at each $z_s$.  \par

Next, we turn our attention to $w \ (p_{\Lambda}=w \rho_{\Lambda})$. Fig.~\ref{fig:omega} shows the difference $\Delta m$ between $m-M$ computed using the {\it WMAP five year}] parameters $w=-1.0$ [blue solid] and that computed by other values; $w=-1.5$ [red solid], $w=-0.5$ [green solid], $w=-1.03$ [green dashed], and $w=-0.99$ [blue dashed]. The deviation $\Delta m(z_s)_{\rm ave}$ and the variance $\sigma_m(z)$ are also placed in the same way as Fig.~\ref{fig:Omega_m}. By analogy to $\Omega_m$, we can see that $w$ has at least the uncertainty $-0.99 \sim -1.03$.

\section{\label{sec6}Conclusions and Discussion}

We have investigated the apparent magnitude in a realistic inhomogeneous universe 
using the general distance formula where  the perturbation of the relative distance 
from the standard FRW distance is expressed by the integral of the matter density contrast along the line of sight.  We used the lognormal PDF to represent the deviation of the apparent magnitude from that in a homogeneous FRW universe with the form 
depending on the averaged mass scale which observed light rays feel along the way. Furthermore, we have considered the transmittance of the light ray, and obtained the deviation of the apparent magnitude for actually observed sources. 
Eventually, we found that the distance contrast $\delta_{d}$ is $\sim 0.01$ for the light rays from sources at $z=2$.  We have used the predicted non-linear power spectrum 
to calculate the correlation function, that is, the variance of the apparent magnitude. 
Finally, using the derived deviation and variance, we have estimated the uncertainty 
in determining the cosmological parameters; $\Omega_m$ and  $w$, which inevitably exist  when we use the magnitude-redshift relation in the observation of SNe Ia.  
The uncertainty is estimated to be  $\Omega_m \sim 0.02$, and 
$w \sim 0.04$.   \par

As we mentioned in Sec.~\ref{sec1}, the variance of the apparent magnitude is also studied in various ways in the many previous works. In particular, in~\cite{2006PhRvD..73b3523B}, the correlation function and the power spectrum of the luminosity distance fluctuations is defined to use them as a new observational tool, which is somewhat similar to our work. In addition, standard lensing convergence analyses (e.g.~\cite{2012ApJ...744L..22S}) make use of the expression which is obtained by integrating matter power spectrum along the line of sight, similar to Eq.~(\ref{am_cor}). However, our work is exactly original and different from these previous studies because we also take account of the effect that collapsed objects are formed when the wave number exceeds the critical value at each redshift, which read to the blocking effect of light rays and the upper limit of the wave number for the matter power spectrum. Thus, our  more realistic scenario may improve the actual observations of convergence.    \par

Future studies will consider the possibility of separating the dispersion of the brightness of an observed source $(m-M)$ into the contribution due to the lensing effect considered in this paper and that due to intrinsic errors associated with the absolute magnitude.  
As mentioned in Sec.~\ref{sec4}, these two effects have different  redshift dependence.  The intrinsic error does not seem to have any redshift dependence.   
Thus, the redshift dependent part of the dispersion may be associated with the effect due to inhomogeneity. Recently, J{\"o}nsson and collaborators have discussed this problem by using dark matter halo models constrained by selected samples of SNe Ia from the first 3 yr of the Supernova Legacy Survey, and found that the lensing dispersion is approximately proportional to the SNe Ia redshift~\cite{2010MNRAS.405..535J}.  Also we can use the {\it three}-point correlation function which indicates  non-Gaussianity, called  {\it skewness}. Since the intrinsic error is considered to be Gaussian~\cite{1999ApJ...517..565P}, we can identify  skewness as the effect of inhomogeneity.  We can predict the skewness by calculating the three-point correlation function using Eq.~(\ref{eq.m-z}) for given cosmological parameters. 
Thus if we could observe many standard candles, e.g. SNe Ia, at same $z$ that we are able to see the skewness as a statistical characteristic of the distribution of their brightness, $m-M$, we can estimate the cosmological parameters by comparing the observed  and  predicted distributions of the apparent magnitude.  \par

We also comment on the choice of  $M_{\rm mim},\ M_{\rm max}$ in Eq.~(\ref{Dm_ave}). 
In Sec.~\ref{sec5}, we set $M_{\rm mim}=10^{10}M_{\odot}, M_{\rm max}=10^{15}M_{\odot}$. Since, from Fig.~\ref{fig:T(z_s;M)Dm(z_s;M)}, we find that the contribution corresponding to  a mass scale exceeding $10^{15}M_{\odot}$ does not exist,  setting  $M_{\rm max}=10^{15}M_{\odot}$ is plausible. On the other hand, in the range $M \lsim 10^{13.5}M_{\odot}$, a smaller  mass scale results in a smaller contribution to the averaged deviation. Hence, if the minimum mass scale $M_{\rm mim}$ is reset as the mass scale of $10^{10}M_{\odot}$ or below, the final result of the averaged deviation 
is not changed so much. 
However, as the mass scale become smaller, the Born approximation, namely the assumption that the path of light rays are  {\it straight} breaks down gradually, and accordingly we can not use the idea of transmittance. Thus, we need to consider a more reasonable setting of the minimum mass scale $M_{\rm mim}$.  \par

Finally, future studies will evaluate the effect of {\it shear} in Eq.~(\ref{eq.d-z}), which has been neglected until now. We need to consider systematic shear because of the matter distribution through which the light rays propagate, but, we can only assume appropriate matter distribution models. We note that we can use the fact that linear density contrast in Eq.~(\ref{del_ave}) is truncated at the upper limit, as cues for  modeling in future studies.

\acknowledgments

We would like to thank A. Taruya for providing us with the code for numerical calculation and N. S. Sugiyama and D. Nitta for useful comments. This work is supported in part by a Grant-in-Aid for Scientific Research from JSPS (No. 26400264  for T.F.).

\bibliographystyle{JHEP.bst}
\bibliography{ms.bib}

\end{document}